\documentclass[12pt]{article}
\usepackage{graphicx}

\def\napoli{Logunov Institute for High Energy Physics, NRC Kurchatov Inst., Protvino, RF}

\def\Title#1{\begin{center} {\Large #1 } \end{center}}
\def\Author#1{\begin{center}{ \sc #1} \end{center}}
\def\Address#1{\begin{center}{ \it #1} \end{center}}

\newenvironment{Abstract}{\begin{quotation}  }{\end{quotation}}
\newenvironment{Presented}{\begin{quotation} \begin{center} 
             PRESENTED AT\end{center}\bigskip 
      \begin{center}\begin{large}}{\end{large}\end{center} \end{quotation}}
\def\Acknowledgements{\bigskip  \bigskip \begin{center} \begin{large}
             \bf ACKNOWLEDGEMENTS \end{large}\end{center}}

\begin{document}
\begin{titlepage}

\vfill
\Title{Diffractive Scattering: Problems in Theory and Praxis}
\vfill
\Author{ Vladimir A. Petrov }
\Address{\napoli}
\vfill
\begin{Abstract}
A concise survey of problems in modern strong interaction physics at high energies is given.
\end{Abstract}
\vfill
\begin{Presented}
Presented at EDS Blois 2017, Prague, Czech Republic, June 26-30,2017
\end{Presented}
\vfill
\end{titlepage}
\def\thefootnote{\fnsymbol{footnote}}
\setcounter{footnote}{0}

\section{Basic principles and their consequences}

Diffractive hadron scattering is normally associated with strong interactions.
The failure of perturbative techniques to describe hadron scattering at the beginning of the 1950s led finally to the rise of the axiomatic method with an accent on the rigorous results following from the general quantum field theory based on 
causality, unitarity, spectrum condition, 
asymptotic completeness, positive definite metric 
of Hilbert space, Poincar\'e invariance without references to interaction Lagrangians \cite{Wightman}. In this framework it was realized that many features of the physical observables stem from general principles independent on specific dynamics. The very relationship of the (unobservable)quantum fields and observable amplitudes(cross-sections)- so-called "reduction formulas" (due to Lehmann-Symanzik-Zimmermann or Bogoliubov- Medvedev – Polivanov) gave a solid ground for model construction. The results of various models (e.g., asymptotic behaviour at high energy) since then must be consistent with the limitations imposed by axiomatic bounds. The very mathematical thoroughness inherent to the axiomatic approach sets a pattern for building comprehensive and falsifiable models.
It would be natural and suggestive to try to check the validity of the very basic principles, e.g. that of micro-causality (field commutators disappear outside the light cone). Normally micro-causality implies analyticity in the momentum space which, in turn, allows to relate, e.g., the real part of the scattering amplitude with its imaginary part with help of so-called dispersion relations(Cauchy theorem adapted to a concrete problem). For instance, we have for the forward scattering amplitude:
\begin{center}
$ Re T(s,0) = P_{N-1}(s) + \frac {(s-s_{0})^{N}} {\pi} \int ds^{'} {ImT(s^{'},0)}/[{(s^{'}-s_{0})^{N} (s^{'}- s)]}.$
\end{center}
 To check this formula we have to know about its constituents. Concerning the integer $ N $ giving the growth power of the amplitude at large $ s $ we know due to Andr\'e Martin that it can't achieve 2, so with $ N=2 $ the integral converges. The position of the "reference point" $ s_{0} $ is arbitrary (outside the cut and possible poles), so we have to have some ideas about the choice of this point allowing to know the number $ P_{0} $. This could be independent symmetry considerations or else but generally it remains arbitrary.Further, the imaginary part is integrated in all interval $ \lbrace threshold, +\infty \rbrace $ which is inaccessible. Now, the real part is to be extracted from the analysis of the data in a narrow interval of Coulomb-nuclear interference.Such an operation is, unfortunately, model-dependent (akin to the extraction of $ x $ and $ y $ from $ \vert x+iy \vert $). So we see that the direct check of the dispersion relations  (and by this the micro-causality)cannot be done in a pure way but should be associated with additional assumptions which different authors can chose according to their personal preferences.
 
Thus, it seems we are to believe John D. Norton \cite{No} :
"...a supposed indispensability or fertility of the notion of causation is at most telling us something about us and does not establish that the world is governed at some fundamental level by a principle of causality."

Is it possible to check experimentally unitarity and other premises of axiomatic QFT?
General answer: "No". But then what can we gain from it? The advantage is
that results of various models (e.g., asymptotic behaviour at high energy) must be consistent-by virtue of our faith in the general principles- with the limitations imposed by axiomatic bounds. This narrows the vast expanse of ideas to reasonable limits. The very mathematical thoroughness and lucidity both in notions and premises inherent to the axiomatic approach sets a pattern for building comprehensive and falsifiable models.

Now something about the high energy bounds.The famous Froissart - Martin bound reads
\begin{center}
$  \sigma_{tot}\leq \frac{\pi}{m_{\pi}^{2}}\ln ^{2}(s/s_{0}).$
\end{center}
If the scale $ s_{0} $ is not specified then the bound is only useful in an abstract sense: if your model gives $ \sigma_{tot}  $ growing like $ \ln ^{2.01}(s/s_{0}) $ at $ s\rightarrow \infty $ then you are possibly in odds with general principles.
Now, much work was spent on obtaining the bounds without arbitrary constants.
In paper \cite{roy} 
there was obtained the upper bound for the "averaged cross-section" for $ \pi\pi $ interaction
\begin{center}
$ \bar{\sigma}_{tot}(s) = \frac{1}{s}\int^{s} d\acute{s} \sigma_{tot}(\acute{s} ).$
\end{center}
which reads
\begin{center}
$ \bar{\sigma}_{tot}(s) \leq \frac{\pi}{m_{\pi}^{2}}[\ln (\frac{s}{s_{0}}) + \frac{1}{2}\ln \ln (\frac{s}{s_{0}})+1]^{2} $
\end{center}
where now the scale $ s_{0} $ is known:
\begin{center}
$s_{0} = \frac{\sqrt{2}m_{\pi}^{2}}{17\pi\sqrt{\pi}} = 2.9\cdot 10^{-4} GeV^{2}. $
\end{center}
Let us see how the bound looks at the LHC energy:
\begin{center}
$ \bar{\sigma}_{tot}(s = (7TeV)^{2}) \leq 51349.82 mb. $
\end{center}
The proton-proton cross-sections as measured by the TOTEM and ATLAS collaborations are  only near 100 mb at 7 TeV. The pion-pion cross-section has to be even less.
So we see that upper bounds derived from the general principles are only useful in an abstract but not in practical sense. They are too high to discriminate model predictions.
Nonetheless we would like to emphasize that axiomatic approach remains to be a useful tool for mathematically and logically consistent formulation of the models.
\section{Models and Theory}
People making models can be divided in two groups. The first group which I call "People of Principle" respect general principles and their consequences and motivate the choice of parameters and trial functions as much as possible.
They try to trace their predictions with the basic assumptions of the model.
They are ready to accept ruling out their models in case of blatant disagreement with experiment.

The second group are "Practical People". Their salient feature is absence of any adherence to any kind of preset dogmas. They take everything which seems useful
and in the case of failure don't hesitate to change parameters and trial functions
without taking the trouble of any motivations.The only criterion is the best chi squared.Their models live a long life permanently changing with any new set of the data. Success and failure are equally insignificant for conceptual conclusions.

Conceptual issues are related to some fundamental basis of the strong interaction phenomena. There is a justified belief that this is QCD. This theory has a remarkable property of the asymptotic freedom which could create an impression that we can control our perturbative calculations at high energy. However, it is only true for so-called "hard" processes related to short distances where strong interactions become actually weak (in the sense of their intensity).
In diffractive processes we deal with a complicated combination of the UV (s-channel)and IR (t-channel)asymptotics.
The latter is related to the plausible "confinement" phenomenon. The general outline of the derivation of the hadronic S-matrix from the generic quark-gluon 
Green function is fairly intelligible. Colourless combinations of quark and/or gluon momenta should be related to a pole at the relevant hadron mass. Multifold residue  at all these poles gives an on-shell hadronic S-matrix.However, practical realisation of such an operation is still unknown.
The fact that diffractive processes are related with long-distance (IR) properties of Yang-Mills fields is implied by the continuous growth with energy both of the transverse and longitudinal sizes of the interaction region. Thus, from the measurements of the forward slope at the LHC ( 7 TeV) we see that the transverse size of the interaction region is of order of 1.25 fermi. The growth is extremely slow: this size is only half as much again as the transverse size at the U-70 Protvino accelerator at $ \sqrt{s} \approx 0.01 TeV$. Meanwhile the longitudinal size grows very rapidly achieving $ \textit{O}(10^{4} $ fm) at the LHC.
So we see that diffraction studies concern the large-distance properties of QCD and this fact relates the models in the field to one of the most challenging problems of modern physics.
To our mind a good model should take as much as possible both from the axiomatic field theory (in what concerns the general features of trial functions) and from QCD (the choice of concrete parameters and, hopefully, of trial functions).
\section{QCD: what we have and what do we wait for?}
It would be naive to hope that we can construct exactly the hadronic S-matrix from the first principles of QCD.The only thing we can hope for is that we would be able to borrow some firm results currently obtained in QCD when constructing models. At present these results are very scarce.

For example, the Pomeron trajectory $ \alpha_{\mathsf{P}} $ at large negative $ t $ degenerates into a weakly perturbed exchange by non-interacting (due to the asymptotic freedom) gluons, i.e.
\begin{center}
$ \alpha_{\mathsf{P}}(t)\rightarrow 1 $(free gluons)$ + \textit{O}(\bar{g}_{s}^{2}(t) $
\end{center}
where QCD effective coupling $ \bar{g}_{s}^{2}(t)\rightarrow 0 $.
Despite the fact that all this applies to small distances, we conclude that, because of the monotony of the Regge trajectory, one of the most fundamental parameters responsible for the behaviour of the total cross-sections, the Pomeron intercept $ \alpha_{\mathsf{P}}(0)$, exceeds 1 sharply. It means that qualitatively QCD predicts the growth of the total cross-sections. However, the calculation of $ \alpha_{\mathsf{P}}(0)$ is still on the agenda  in spite of great efforts spent during several decades. The problem  is all the more difficult since the Pomeron intercept (at least in the commonly used approximation of pure gluodynamics) does not depend on the QCD coupling constant.

Another important parameter is the slope of the Pomeron, $\alpha_{\mathsf{P}}'(0) $. Both from the rigorous consideration and common physical sense it follows that 
\begin{center}
$ \alpha_{\mathsf{P}}'(0)\sim \Lambda_{QCD}^{-2} $
\end{center}
which means that this quantity is highly non-perturbative because
\begin{center}
$\Lambda_{QCD}^{-2}\sim \exp (1/\beta_{0}g_{s}^{2}) . $
\end{center}
  But if we even know the value of $ \Lambda_{QCD} $ there remains theoretical uncertainty because of the renormalization scheme dependence.
  Thus at the moment we have no firmly established results from QCD which could be used in constructing diffraction scattering models.
  
 \section{Are we in a "truly asymptotic regime"?} 
 
 Prior to try to answer the question we have to precise what do we mean under
 "truly asymptotic regime". Generally this is related to a belief that at some "high enough energies" the energy evolution of such basic quantities as total and elastic cross-sections, forward slopes of the differential cross-sections, sometimes the very differential cross-sections acquires a stable character with some simple functional dependences or even the energy independence.
Some extreme motto was advocated in the 1960s by G. Chew and S. Frautschi :"Strong interactions should be as strong as possible" and  "truly asymptotic regime" would mean the saturation of the upper bounds. 
 However, we have seen in the beginning of my talk that the rigorous upper bounds for the total cross-sections are more than five hundred times higher than the values of the LHC measurements.
 We also could try the upper bound for the forward slope 
 
 \begin{center}
$ B(s)\equiv \left[{\frac{\partial}{\partial t}\left(\frac{d\sigma}{dt}\right)}/({\frac{d\sigma}{dt}}\right)](s, t =0).$
\end{center}
 
 The upper bound for this quantity was derived in \cite{roy 2} and reads
\begin{center}
$ B(s)\leq \frac{1}{8m_{\pi}^{2}} \ln^2 \frac{s}{s_{0}^{2}\sigma_{tot}} $ 
\end{center}
at $s \gg s_{0}$ .
 If we assume $ s_{0} = 100 GeV^{2} $ then at the LHC ( with $ s = 5\cdot 10^{5} s_{0} $) we have, taking use of the TOTEM data on the total cross-section, 
\begin{center}
$ B(s)\leq 56.8 GeV^{-2} $ 
\end{center}
while the TOTEM measurements \cite{T1} give 
\begin{center}
$ B(s)\approx 20 GeV^{-2} .$ 
\end{center}
This disagreement is not so discouraging as that for $ \sigma_{tot} $ but nonetheless shows that we are far from saturation in this case either.

Generic Regge-eikonal models give the following estimate for the asymptotic ratio
\begin{center}
$  \frac{\sigma_{tot}}{B}\mid_{asymptotic} = 8\pi \approx 25 $
\end{center}
to compare with
\begin{center}
$  \frac{\sigma_{tot}}{B}\mid_{LHC} = 12.5$

\end{center}
These estimates show clearly that even with the highest LHC energy we are still far from a stably sustained regime of the "black disc" type.

Now, something about noticeable craze of use (and, probably, even misuse!) of a " $ \ln ^{2} s $ paradigm" for modelling $ \sigma_{tot} $ and $ B $. It is evident that this $ \ln ^{2}$ is induced by the Froissart bound as a sign of "asymptoticity". Moreover, the function is seductively simple and qualitatively well corresponds to the observed slow growth of the cross-sections.
On the other hand , Regge-eikonal models lead indeed to such a behaviour for $ \sigma_{tot} $ and $ B $. But at which energies?
A reasonable criterion of the asymptotic behaviour ( or, at least, one of its important signatures)is the condition that strong interaction bears a universal character, independent of a specific flavour content of the colliding hadrons.
One of the important charaсteristics is the proper size of hadrons related to their valence cores. For instance, the proper size of the nucleon $ \langle r^{2}\rangle_{nucleon}^{1/2} $ can be extracted from the proton and neutron form factors and reads \cite{Pe} 
\begin{center}
$ \langle r^{2}\rangle_{nucleon} = r^{2}_{e}(proton) + r^{2}_{e}(neutron) = 0.654\pm 0.002 fm^{2} = ( 0.809 \pm 0.007 fm)^{2} $
\end{center}
In the impact parameter plane it is "seen" as
\begin{center}
$\langle b^{2}\rangle_{nucleon} = (0.660\pm 0.007 fm)^2 = 11.197\pm 0.157 GeV^{-2}  $ 
\end{center}
The transverse extent of the interaction region is given by the magnitude $B$.
So we are in asymptotic region if 
\begin{center}
$B_{pp}\gg \langle b^{2}\rangle_{nucleon} . $
\end{center}
The available estimates show that even for the ratio $ B_{pp}/ \langle b^{2}\rangle_{nucleon} $ to be equal to 3 we would need the energy of the order of $ 10^{3}\div 10^{4} TeV $. Nothing to say about $ \gg $ !
So, in no way the specific features of the colliding hadrons may be already completely neglected while the cross-sections still bear information about their parton structure.
\section{Conclusions} 
The matter of fact is that  - in a sharp distinction with the EW part of the SM - the strong interaction part is in a very pitiful state having no regular and firm predictions from its  fundamental basis, i.e. Quantum Chromodynamics, which could give us an opportunity to subject it to scrutiny in the kinematical region where quark-gluon interactions are genuinely strong.
We have only scarce indications to the qualitative agreement of the  observed hadron scattering with general premises of QCD. Whence comes an urgent need in the solid theoretical progress in studies of QCD at large distances/high energies.
Models using mostly terminological relation with QCD can hardly help. Up to the present time no one model can -even in the case of successful description of the data - give us a clear and profound insight in the mechanism of the diffractive scattering. Thoroughness in the model formulation (with axiomatic approach as a specimen) would be very helpful in discrimination of their heuristic value.
Experiments give us more and more valuable material for description and explanation. However, the former grossly prevails over the latter.
Nonetheless, the very observed trends imply quite interesting conclusions.
On one hand, we see the growth of the interaction region with energy while on the other this growth is extremely slow.
 This fact shows that in spite of quite a high collision energy we are far from "ASYMPTOPIA" but "in return" we still can explore many interesting features of hadronic structure in the realm of diffractive scattering. In a sense we are in a borderland between ondulatory and corpuscular faces of high energy phenomena which makes diffraction scattering extremely important and interesting field of modern physics.

\Acknowledgements
I am grateful to organizers of this workshop for invitation and creative atmosphere of the meeting.


\begin{thebibliography}{99}


\bibitem{Wightman}
A. S. Wightman, 
Phys.Rev. {\bf 101}, 860 (1956);


N. N. Bogoliubov, 
The talk at the International Congress on Theoretical Physics.
Seattle(USA), 19 September 1956.

Comprehensive exposition of the basics and methods of the axiomatic approach can be found in

R.F. Streater, A.S. Wightman, 

PCT, Spin and Statistics, and All That.Benjamin, New York ( 1964);

R. Jost, 
The General Theory of Quantized Fields. 

Am.Math.Soc., Providence,Rhode Island (1965);


N. N. Bogoliubov, A. A. Logunov, A. I. Oksak, I. T. Todorov,  

General Principles of Quantum Field Theory. 

Dordrecht [Holland]; Boston, Kluwer Academic Publishers(1990) 

\bibitem{No}
J.D. Norton, "Causation as Folk Science" Philosophers' Imprint Vol. 3, No. 4.p.7
(2003)

\bibitem{roy}
 	
Andr\'e Martin, S.M. Roy, Phys.Rev. \textbf{D91}, no.7, 076006 (2015)
 
\bibitem{roy 2}
S. M. Roy, Phys. Rep.\textbf{5C}, 176 (1972)

\bibitem{T1} The TOTEM Collaboration, EPL \textbf{101}, 21004( 2013)

\bibitem{Pe} V. A. Petrov, EPJ Web Conf. \textbf{138},02008 (2017) 


\end{thebibliography}
\end{document}